\documentclass[12pt]{article}
\usepackage{epsfig}
\usepackage{rotate}

\begin{document}
%\draft
%\tighten \vspace{5cm}
\title{The Dark Energy in Scalar-tensor Cosmology}
%\vspace{2cm}
\author{Mian WANG \\
 Department of Physics, Henan Normal University \\
 Xinxiang, Henan, China  453002\\
  e-mail: wangm@henannu.edu.cn }
%\date{}
\def\baselinestretch{1.5}
\maketitle
%\begin{document}
\def\be{\begin{equation}}
\def\ee{\end{equation}}
\def\vp{\vspace{3 mm}}
\def\f{\frac}
\vspace{2cm}
\begin{abstract}
Recent observations confirm that our universe  is flat and
consists of a dark energy component with negative pressure. This
dark energy is responsible for the recent cosmic acceleration as
well as determines the feature of future evolution of the
universe. In this paper, we discuss the dark energy of the
universe in the framework of scalar-tensor cosmology. In the very
early universe, the gravitational scalar field $\phi$ plays the
roll of the inflaton field and drives the universe to expand
exponentially. In this period the field $\phi$ acts as a
cosmological constant and dominates the energy budget, the
equation of state ( EoS ) is $w=-1$. The universe exits from
inflation gracefully and with no reheating. Afterwards, the  field
$\phi$ appears as a cold dark matter and continues to dominate the
energy budget, the universe expands according to $\frac{2}{3}$
power law, the EoS is $w=0$. Eventually, by the epoch of $z\sim
O(1)$, the field $\phi$ contributes a significant component of
dark energy with negative pressure and accellerates the late
universe. In the future the universe will expand
acceleratedly according to  $a(t)\sim t^{1.31}$.\\
\\
PACS number(s): 98.80.Cq, 04.50.+h
\end{abstract}
\vspace{3cm}
\newpage
Recent observations of large scale structure, the Hubble diagram
of type Ia supernovae and the angular power spectrum of the cosmic
microwave background {\sl etc }, all indicate that the universe is
flat and is undergoing cosmic acceleration by virtue of  a dark
energy component with negative pressure \cite{R}-\cite{TP}. The
simplest model of the dark energy may be Einstein's cosmological
constant \cite{We}-\cite{PP}.  The another paradgim is the
quintessence \cite{W}-\cite{Z} in the form of a scalar field
slowly evolving down a potential with negative pressure, scenarios
of which are built with dilaton \cite{di}-\cite{Be}, axion
\cite{ax}-\cite{DBS} , tachyon \cite{ta}-\cite{Gi} or other exotic
fields such as phantom \cite{ph}-\cite{Ck}, chaplygin gas
\cite{ch}-\cite{Ss} and so forth \cite{PP}-\cite{Ks}. It is of
fundamental importance to investigate the nature of this energy
component, because it is not only responsible for the present
accelerated expansion but also will determine the fate of the
universe.  For example, if it is a cosmological constant
$\Lambda$, the universe will expand exponentially in the future.
On the other hand, if the dark energy is the energy of a slowly
changing scalar field $\varphi$ with $w_\varphi\simeq -1$ and the
potential $V(\varphi)$ driving the present stage of acceleration
decreases slowly and vanishes eventually, the speed of expansion
decreases after a transient de Sitter-like stage and reaches
Minkowski regime \cite{W}, the universe becomes increasingly cold
and empty. Or the potential falls to $V(\varphi)=-\infty$, the
universe eventually collapse, even if it is flat \cite{Felder}. As
for the case of phantom energy, in which $w<-1$, the sum of the
pressure and energy density is negative. The positive
phontom-energy density becomes infinite in finite time, overcoming
all other forms of matter, such that the gravitational repulsion
rapidly brings our brief epoch of cosmic structure to a close. The
phontom energy rips apart the Milky Way, solar system, Earth, and
ultimately the molecules , atoms, nucleons of which we are
composed, before the death of the universe in a " Big-Rip "
\cite{Ck}. In this letter we investigate the dark energy in the
framework of scalar-tensor cosmology \cite{Wang} with an added
potential term. We will show that the dark energy is from the
gravitational scalar field.  In the future the universe will
expand acceleratedly according to $a(t)\sim t^{1.31}$ with a
constant expansion index. However, the matter density parameter
$\Omega_m$ remains $0.024$, the total mass of a given comoving
volume increases  as $\sim t^2$. The world is not cold and empty,
new cosmic structures are being formed steadily.\\

Scalar-tensor theories of gravity are the most natural
alternatives to Einstein's general relativity. The gravitational
fields are enlarged by introducing a gravitational scalar field.
In what concerns inflationary cosmology, scalar field has
fundamental significance: the inflaton must be scalar field, the
dark energy has to be owed from scalar field. In our model, the
gravitational scalar field $\phi$ plays the role of inflaton and
contributes dark matter and dark energy in succeeding cosmic
evolution stages. Furthermore, scalar-tensor cosmology has some
resemble features with string cosmology, e.g. the  field $\phi$
may be identified with the dilaton field when it is
scaled.\\

\noindent {\bf The scalar-tensor cosmology}\ \ Five years ago, we
have proposed a model of scalar-tensor cosmology \cite{Wang}. In
this model, the motion of  field $\phi$ is described by a
double-well potential $V(\phi)=\beta (1-\phi^2)^2$ equipped by the
coupling function $\omega(\phi)$ and the cosmological function
$\lambda(\phi)$\footnote{Neglect the $U(\phi)$ term, the
derivative of the potential of the motion of $\phi$ is given by
 $V'(\phi)=(2\phi^2\lambda'-2\phi\lambda)/(2\omega+3)=-4\beta\phi
 (1-\phi^2)$, integrating gives  $V(\phi)=\beta (1-\phi^2)^2$. }.
The universe is created when $\phi$ emerges from its tunnelling at
the point $\phi=0$, then begins to inflate exponentially when
$\phi$ rolls from the top-hill $\phi=0$ towards the potential
bottom $\phi=1$. The field $\phi$ plays the role of inflaton field
and acts as a cosmological constant with  EoS $w=-1$. Now, to take
account of the dark energy component, we assume a modified action
of \cite{Wang} by adding a $U(\phi)$ term, it is as follows
 \be
   {\cal A}=\int d^4x \sqrt{-g}\,\,\left[-\phi R
    - \frac{\omega}{\phi} \phi\cdot\phi -2 \phi \lambda(\phi)
   -2 \phi U(\phi)
      -\frac{\Gamma (u\cdot\phi)^2}{1-\phi}+16\pi L_m\right] ,
   \label{fb1}
 \ee
 in which $L_m$ is the Lagrangian of matter,
$\phi\cdot\phi\equiv\phi_{,\sigma} \phi^{,\sigma}$,
 $u\cdot\phi\equiv u_{\mu}\phi^{,\mu}$,
 $u_{\mu}$ is the four-velocity, $\phi^{,\mu}\equiv
\partial\phi/\partial x_{\mu}$, and $\Gamma$ is a constant of
 mass dimension 0. The $\Gamma-$term describes the coupling of the
 field $\phi$ with matter which is described as an ideal fluid.
The coupling function $\omega(\phi)$ and the cosmological function
$\lambda(\phi)$ are proposed to be
 \be 2\omega(\phi)+3 =
\frac{\xi}{1-\phi}\, , \label{fb2} \ee \be \lambda(\phi) = 2\xi
\beta (1-\phi- \phi \,\, ln\,\phi) , \label{fb3}
 \ee
where $\xi$ and $\beta$ are two dimensionless constants, here we
set $\xi=7.5,\,\,\beta=1.7\times 10^{-16}$  and  $\Gamma=0.066$.
The  potential $U(\phi)$ is proposed to be \be
        U(\phi)=\beta^2\sqrt{1-\phi} \; \; e^{-\alpha\phi}\label{fb4}
\ee in which $\alpha$ is a constant of order $O(10^2)$, it is set
to be $81.82$ in the model in order to let the
dark energy coincide with the maybe cosmological constant. \\

Then the Friedmann equations read \be
H^2=\frac{8\pi}{3\phi}(\rho_{m}+\rho_\phi)-\frac{k}{a^2} \, ,
\label{fb5}
 \ee
\be
 \frac{\ddot a}{a}=-\frac{4\pi}{3\phi}(\rho_{m}+\rho_\phi+ 3p_m+3p_\phi)\, ,
\label{fb6}
 \ee
and equations of motion of $\phi$ and $ \rho_m $ are
\begin{eqnarray}
\ddot{\phi} + 3H \dot{\phi} &+& \frac{\dot{\phi}^2}{2 (1-\phi)}
 + \frac{8\pi(3p_m-\rho_m)}{2\omega+3}-\frac{\Gamma}{2\omega+3}
 \frac{\phi\dot\phi^2}{(1-\phi)^2}
 -\frac{2\Gamma}{2\omega+3}\frac{\phi\ddot\phi}{1-\phi} \nonumber \\
 &=&4\beta \phi (1-\phi^2)+\frac{2}{\xi}\phi\sqrt{1-\phi} \;\beta^2 \; e^{-\alpha\phi}
 \left(1-\frac{1}{2}\phi+\alpha(1-\phi)\right),
\label{fb7}
\end{eqnarray}
\be \dot{\rho}_m + 3 H(\rho_m + p_m)
=\frac{\Gamma}{4\pi}\,\left(\frac{\dot\phi\ddot\phi}{1-\phi}
+\frac{1}{2}\frac{\dot{\phi}^3}{(1-\phi)^2}+3H\frac{\dot\phi^2}{1-\phi}
\right) , \label{fb8}
\ee where $\rho_{m}$ , $p_m$ and $\rho_\phi$
, $p_\phi$ are energy density and pressure of the matter  and
field $\phi$
 respectively. The expressions are
  \be
\rho_{\phi}=\frac{1}{8\pi} \left(\phi \lambda + \phi U
  +\frac{\omega
  \dot{\phi}^2}{2\phi}-3 H \dot{\phi}
  -\frac{3}{2}\frac{\Gamma\dot{\phi}^2}{1-\phi}\right) , \label{fb9}
\ee
 \be p_{\phi} = \frac{1}{8\pi} \left(-\phi \lambda -\phi U+
\frac{\omega\dot{\phi}^2}{2\phi}+2 H \dot{\phi}+\ddot{\phi}
-\frac{1}{2}\frac{\Gamma\dot{\phi}^2}{1-\phi}\right) .
\label{fb10} \ee
 These basic equations determine the dynamic evolution of our
 universe.\\

\noindent {\bf Exponential inflation}\ \  When the field $\phi$
rolls down the potential hill from $\phi\simeq 0$, the universe
expands exponentially as described in \cite{Wang} since $U(\phi)$
can be neglected when $\phi<1$, as it is of order $O(\beta)\sim
O(10^{-16})$ compared to $\lambda(\phi)$. The Hubble parameter $H$
is determined by the value of $\lambda(\phi)\simeq 2 \xi\beta$
when $\phi\ll 1$. In the duration of inflation, the energy density
of the scalar field $\rho_\phi$ dominates and $p_\phi=-\rho_\phi$.
It is evident that $\lambda(\phi)$ acts as a cosmological
constant.
The solutions are the same as in \cite{Wang}:\\
\be
       a(t)=a_{0}\,e^{H(t-t_0)} , \label{inf1}\ee
\be
   H=\sqrt{2\xi_e\beta/3}\; ,\ \ \  \ \ \  \xi_e=\xi-1.6\; ,
\label{inf2} \ee
   \be \phi(t)=\phi_0e^{D(t-t_0)}, \label{inf3}
\ee
and \be
      \rho_m=\frac{\Gamma D^2 (3H+D)}{4\pi (4H+2D)}\phi^2(t),
          \label{inf4} \ee
where \be
              D=(\sqrt{1+8/(3\xi_e)}-1)\sqrt{3\xi_e\beta/2}\; .
                       \label{inf5} \ee

 We see that the universe expands exponentially in this inflation
period, the EoS is $w=-1$.   As $\phi$ increases near to $1$, the
$\dot\phi^2$ term in equation (\ref{fb9}) and (\ref{fb10})
increases also, therefore $w_\phi\equiv p_\phi/\rho_\phi$
increases quickly. The inflation ends dynamically when
$\rho+3p\simeq \rho_\phi+ 3p_\phi$ increases from $-2\rho_\phi$ to
zero ($0.9<\phi<1$). It is comfortable that the inflation has
graceful exit. Furthermore, at this epoch, the density of matter
$\rho_m \sim  10^{-19}$ \footnote {This is calculated from
equation(\ref{inf4}) with $\phi\sim 1$.} and the temperature is
$10^{14}$ GeV, so this cosmology does not require reheating and
does not suffer the monopole problem, it is hot enough to match
the temperature of Cosmic Microwave Background $(2.73 K)$ but
lower than the GUT scale $10^{16}$ GeV. After inflation, the
universe enters  the era
of power-law expansion.\\

\noindent {\bf Power-law expansion}\ \ After inflation, the energy
density $\rho_\phi$ remains to be dominating  and the scalar
$\phi$ behaves as massive particle with effective mass $\sim
2\sqrt{\beta}$ when $\phi$ goes near to the bottom of potential
well ($\phi=1$). That is to say, the energy density of the scalar
field $\rho_\phi$ behaves as  cold dark matter, then the universe
experiences  power-law expansion.

  Let
$$
   \phi=1-\sigma^2,
$$
 it follows that
\be
   (1-2\gamma)\ddot{\sigma}+3H\dot{\sigma}+4\beta\sigma=
    -sign(\sigma)\frac{\beta^2e^{-\alpha}}{2\xi},\label{fb11}
\ee
 in which
$$
        \gamma\equiv\frac{\Gamma}{\xi}.
$$
  From equation (\ref{fb11}), we see that the $\sigma(t)$ will
oscillate when the damping $\frac{2}{3}H$ decreases to be less
than the critical damping, i.e., $H<\frac{4}{3}\beta$. This phase
transition is also connected with the energy condition
$\rho+3p=0$. Then we write

\be \sigma(t)=f(t)\;\cos\,\omega_1 t,\label{w1} \ee
we obtain \be
         \omega_1^2= \frac{4\beta+2\beta^2e^{-\alpha}/\pi\xi
         f}{1-2\gamma},
       \label{w2}\ee
and equations:
 \be
            \dot f =-\frac{3H}{2(1-2\gamma)}f, \label{fb12}
\ee \be
          H^2=\frac{4\xi\beta}{3} f^2+
               \frac{\beta^2e^{-\alpha}}{\pi}f
               + \frac{8\pi}{3}\rho_m ,\label{fb13}
   \ee
and
 \be
      \dot{\rho}_m + 3 H(\rho_m + p_m)
      =\frac{\Gamma}{\pi}\,\left(\frac{3\beta^2e^{-\alpha}}{2\pi\xi}
       H f+3\beta Hf^2 \right) . \label{fb14}
\ee
 The energy density of $\phi$ is:
\be
           \rho_{\phi}= \frac{\xi\beta}{2\pi} f^2 +
              \frac{3\beta^2e^{-\alpha}}{8\pi^2}f,  \label{fb15}
\ee and the pressure of $\phi$ is
\be
        p_\phi=-\frac{\beta^2e^{-\alpha}}{8\pi^2}f. \label{fb16}
\ee

These equations are the smoothed Friedmann equations and equations
of motion of $\phi$ and $\rho_m$ in case of oscillating $\phi$,
that is, quantities of which are mean values in time (over a
period). It is in need to notice that the EoS can't be written as
$<p>/<\rho>$. It should be $w=<p/\rho>$ in the case of oscillating
quantities.\\

 We notice that in equation(\ref{fb15}), the multiplier constants
 $\frac{\xi\beta}{2\pi}=O(10^{-16})$ and
 $\frac{\beta^2e^{-\alpha}}{4\pi^2}=O(10^{-69})$. Therefore, when
 $f>10^{-53}$, $ p_\phi\simeq 0$, the dark matter dominates,
 equation(\ref{fb12}) has solutions as \cite{Wang}
 \be
               f=\frac{1}{\sqrt{3\xi\beta}t}, \label{fb17}
\ee  \be
          H=\frac{2(1-2\gamma)}{3t},\ \ \ \ \ \ a(t)\sim t^
          {\frac{2}{3}(1-2\gamma)}, \label{fb18}
\ee and \be
          \rho_m=\frac{(1-2\gamma)\gamma}{\pi t^2}.
\ee Clearly the cosmic scale $a(t)$ expands according to
$\frac{2}{3}$-power law nearly. The EoS is $w=0$. In this period
the total mass in a given co-moving volume $\sim \rho_m(t) a(t)^3$
is nearly constant.\\

\noindent {\bf Deceleration-acceleration transition }\ \
 When $f(t)$ decreases to be less than $10^{-53}$, corresponding to
$t>10^{59}\, M_P^{-1}$, the dark energy component becomes
important gradually. In this transient stage, we do not have
analytic solutions. Numerical calculation shows that
the deceleration-acceleration transition occurs at $z=0.51$. \\

For the present universe, we adopt \cite{S} $H_0=71\, km/sMpc=
1.24\times 10^{-61}\, M_P$, from numerical data we determine the
age of universe $t_0=8.50\times 10^{60} M_P^{-1}=14.5\, Gy$, and
we have
  $\Omega_{\rho_m}=0.04$.  It is
found that\footnote{For $a(t)\sim t^n$, there is $H\equiv
\frac{\dot a}{a}=\frac{n}{t}$.}
 \be
    H(t)=\frac{1.06}{t},\ \ \ \ \ \  a(t) \sim t^{1.06}, \label{fb19}
\ee
  The EoS  is
 $w=-0.37$ \footnote{By $n=\frac{2}{3(1+w)}$.}  corresponding to
  $ a(t) \sim t^{1.06}$.\\

\noindent {\bf The future evolution of the universe} \ \ After
this transition period we can obtain an analytic solution. In
fact, when $t>>10^{60}\, M_P^{-1}$, the dark energy varies as
$\sim t^{-2}$. From the equation(\ref{fb15}), we have $f(t)\sim
t^{-2}$, hence  equation(\ref{fb12}) have solutions \be
    H(t)=\frac{4(1-2\gamma)}{3t}=\frac{1.31}{t}, \label{fb20}
\ee\be
    a(t)=a(t_1) \left(\frac{t}{t_1}\right)^{1.31},  \label{fb21}
\ee
and \be
    f(t)=f(t_1) \left(\frac{t_1}{t}\right)^2,\ \ \ \ \ \  \rho_m(t)=\rho_m(t_1)
    \left(\frac{t_1}{t}\right)^2 , \label{fb22}
\ee for $t_1>>t_0$. The future $\Omega_m$ is a constant $0.024$.
 We see that $a(t)\sim t^{1.31}$ and the
EoS  is $w=-0.491$ correspondingly. For clearness, we plot the EoS
parameter $w$ as function of  redshift, $z$, in Fig. 1 in which we
see that $w$ decreases with the decrease of $z$ and approaches
$-0.491$.
\begin{figure}[htbp]
\includegraphics[angle=-90, width=1.0\textwidth]{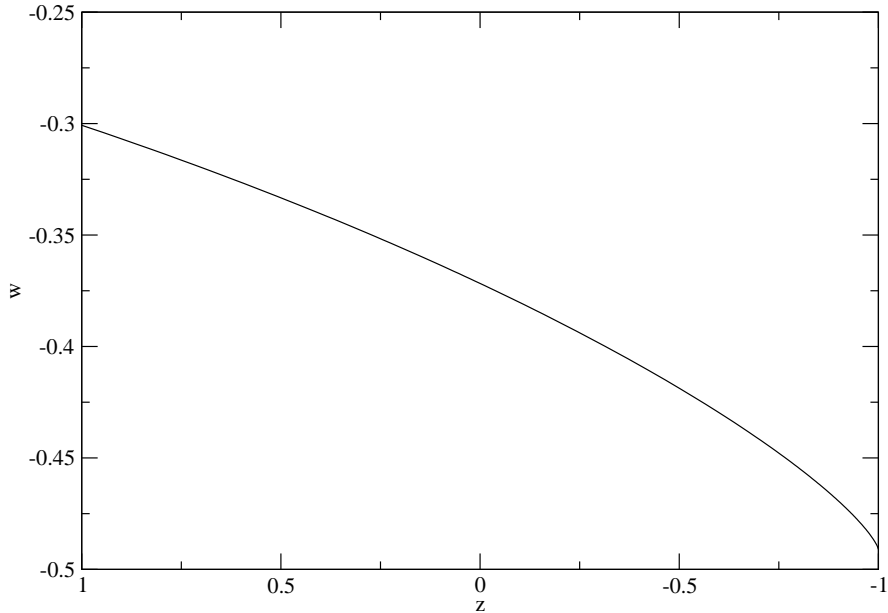}
\caption{\ \ \ w\,=\,-\,0.334 \ at \ z\,=\,0.51. }
\end{figure}

The total mass in a given co-moving volume varies as $\sim
\rho_m(t) a(t)^3 \sim t^2 $. That is to say, in the future
inflationary universe, matter is created steadily, new stars and
galaxies will form
succeedingly. \\

\noindent {\bf Summary and discussions.}\ \  We have shown that in
the early universe   the gravitational scalar field  $\phi$ plays
the role of inflaton field driving a period of exponential
inflation. The exit from inflation is carried out dynamically via
the transition of $p_\phi-3\rho_\phi$ from negative value to
positive. In this inflation stage, the $\lambda(\phi)$ acts as a
large
cosmological constant.\\

  After inflation of the early universe
the scalar field  $\phi$ behaves as a cold dark matter, the
universe expands according to $\frac{2}{3}$ power law \footnote{
In this period, $\Omega_m =0.05$ and $\Omega_\phi=0.95$.}.  After
the time $10^{60}\, M_p^{-1}$, the dark energy contributed by
field $\phi$ increased significantly and begins  to accelerate the
universe at $z=0.51$. Eventually our universe will expand with a
power law $a(t)\sim t^{1.31}$, it is spatially flat although it
is closed.\\

It is manifest that  the gravitational scalar field $\phi$
attributes a varying cosmological constant decreasing as $\sim
\xi(t) \sim t^{-2}$ after the deceleration-acceleration
transition.\\

Finally, it is worthwhile to point that the universe is created
from the instanton solution \cite{Wang} of field $\phi$ when field
$\phi$ is dropped in the left potential well for negative
$\phi$.\footnote{ When $\phi$ is negative in relation to the
problem of the birth of our universe, the function $U(\phi)$
should be written as $ U(\phi)=\theta(\phi)\beta^2\sqrt{1-\phi} \;
e^{-\alpha\phi}$, where $\theta(\phi)$ is the step function:
$\theta(\phi)=0 $ for $\phi<0$ and $\theta(\phi)=1$ for $\phi>0$.}
This instanton associates with the condition $k=1$ provides the
birth of a closed universe rather than the spatially flat FLRW
universe or the Big Bang. Yet after a few Hubble times, inflation
makes the curvature $1/ a^2$ can be neglected, so $\Omega\sim 1$.
That is to say, we have  $\Omega\sim 1$ without fine-tune.


\begin{thebibliography}{99}
\bibitem{R}
 A.G. Riess {\sl et al.,}  Astron. J. {\bf116} (1998) 1009.
\bibitem{P}
 S. Perlmutter et al., Astrophys. J. {\bf517} (1999) 565.
\bibitem{CP}
 C. Pryke {\sl et al.,} Astrophys. J. {\bf 568} (2002) 46.
\bibitem{V}
 L. Verde {\sl et al.,} MNRAS {\bf 335} (2002) 432.
\bibitem{CA}
 R.R. Caldwell, R. Dave and P.J. Steinhardt,  Phys. Rev. Lett. {\bf 80}
 (1998)1582
\bibitem{B}
 N. Bahcall {\sl et al.,} Science {\bf 284} (1999) 1481.
\bibitem{S}
 D.N. Spergel {\sl et al.,} Astrophys. J. Suppl. {\bf 148} (2003)175.
\bibitem{JA}
 H.K. Jassal, J.S. Bagla and T. Padmanabham, Phys. Rev. D {\bf 72}
 (2005) 103503.
\bibitem{U}
 A. Upadhye, M. Ishak and  P.J. Steinhardt,  Phys. Rev. D {\bf 72}
 (2005) 063503.
\bibitem{OS}
 J.P. Ostriker and P.J. Steinhardt, Science {\bf 300} (2003) 1909.
\bibitem{TP}
 T. Padmanabham, Curr. Sci. {\bf 88} (2005) 1057.
\bibitem{W}
 C. Wetterich, Nucl. Phys. B {\bf 302} (1988) 668;\ \ \ \ Astron.Astrphys. {\bf
 301} (1995) 321; \ \ \ \  Phys.Lett. B{\bf 497} (2001) 281;
 \ \ \ \  Phys.Lett. B{\bf 528} (2002) 175;
 \ \ \ \    Phys.Lett. B{\bf 594} (2004) 17.

\bibitem{We}
 C. Wetterich, Phys. Rev. Lett. {\bf 90} (2003) 231302.
\bibitem{MA}
 M.K. Mak and T. Harko, Int. J. Mod. Phys.  D {\bf 11} (2002) 1389.
\bibitem{RR}
 R.R. Caldwell {\sl et al.,} ApJ. {\bf 591} (2003) L75.
\bibitem{RE}
 R.R. Caldwell and E.V. Linder,  Phys. Rev. Lett. {\bf 95} (2005) 141301.
\bibitem{Z}
 I. Zlatev, L. Wang and  P.J. Steinhardt, Phys. Rev. Lett. {\bf 82} (1999) 896.
\bibitem{di}
 R. Bean and J. Magueijo,  Phys.Lett. B{\bf 517} (2001) 177.
\bibitem{LA}
 L. Amendola  {\sl et al.,} Phys. Rev.  D {\bf 67} (2003) 043512.
\bibitem{Su}
 M. Susperregi,  Phys. Rev.  D {\bf 68} (2003) 123509.
\bibitem{Be}
 E.A. Bergshoeff  {\sl et al.,} Class. Quant. Grav.  {\bf 22} (2005) 4763.
\bibitem{ax}
 P. Jain,  Mod. Phys. Lett. A {\bf 20} (2005) 1763.
\bibitem{Mb}
 R. Mainini and S.A. Bonometto,  Phys. Rev. Lett. {\bf 93} (2004) 121301.
\bibitem{DBS}
 S.M. Barr and D. Seckel,  Phys. Rev.  D {\bf 64} (2001) 123513.
\bibitem{ta}
  J.S. Bagla, H.K. Jassal and T. Padmanabham, Phys. Rev. D {\bf 67}
 (2003) 063504.
\bibitem{Co}
 E.J. Copeland {\sl et al.,} Phys. Rev. D {\bf 71} (2005) 043003.
\bibitem{Gi}
 G.W. Gibbons,  Class. Quant. Grav.  {\bf 20} (2003) S321.
\bibitem{ph}
  J.G. Hao and X.Z. Li,  Phys. Rev. D {\bf 67} (2003) 107303;
\bibitem{Ha}
  J.G. Hao and X.Z. Li,   Phys. Rev. D {\bf 70} (2004) 043529.
\bibitem{Ck}
 R.R. Caldwell, M.
  Kamionkowski and N.N. Weinberg,  Phys. Rev. Lett. {\bf 91} (2003) 071301.
\bibitem{ch}
  M.K. Mak and T. Harko, Phys. Rev.  D {\bf 71} (2005) 104022.
\bibitem{Bs}
 T. Barreiro and A.A. Sen, Phys. Rev.  D {\bf 70} (2004) 063511.
\bibitem{Ss}
 A.A. Sen and R.J. Scherrer,  Phys. Rev.  D {\bf 72} (2005) 104022.
\bibitem{PP}
 P.J.E. Peebles, Rev. Mod. Phys. {\bf 75} (2003) 559.
\bibitem{Ks}
 A. Kusenko and P.J. Steinhardt, Phys. Rev. Lett. {\bf 87} (2001) 141301.
\bibitem{Felder}
 G. Felder, A. Frolov, L. Kofman and A. Linde,  Phys. Rev. D {\bf 66} (2002) 023507.
\bibitem{Wang}
 M. WANG, Phys. Rev. D {\bf 61} (2000) 123511.
\end{thebibliography}
\end{document}